\documentclass[]{spie}  

\usepackage{amsmath,amsfonts,amssymb}
\usepackage[colorlinks=true, allcolors=blue]{hyperref}
\usepackage{textcomp}
\usepackage[euler]{textgreek}
\usepackage{tabularx}
\usepackage{caption}
\usepackage{subcaption}
\usepackage{graphicx}
\usepackage{xcolor}

\usepackage[normalem]{ulem}

\newcommand{\angstrom}{\mbox{\normalfont\AA}}

\title{\scalebox{2}\textmugreek-Spec Spectrometers for the EXCLAIM Instrument  }

\author[a,b]{Mona Mirzaei}
\author[a]{Emily M. Barrentine}
\author[a]{Berhanu T. Bulcha}
\author[a]{Giuseppe Cataldo}
\author[d]{Jake A. Connors}
\author[a]{Negar Ehsan}
\author[a]{Thomas M. Essinger-Hileman}
\author[a]{Larry A. Hess}
\author[a,e]{Jonas W. Mugge-Durum}
\author[a]{Omid Noroozian}
\author[a,f]{Trevor M. Oxholm}
\author[a]{Thomas R. Stevenson}
\author[a]{Eric R. Switzer}
\author[a,e]{Carolyn G. Volpert}
\author[a]{Edward J. Wollack}
\affil[a]{NASA Goddard Space Flight Center, Greenbelt, MD, USA}
\affil[b]{Science Systems and Applications, Inc., Lanham, MD, USA}
\affil[c]{University of Maryland, Baltimore County, MD, USA}
\affil[d]{National Institute of Standards and Technology, Boulder, CO, USA}
\affil[e]{University of Maryland, College Park, MD, United States}
\affil[f]{University of Wisconsin, Madison, WI, United States}

\authorinfo{Further author information: (Send correspondence to M.M. and E.M.B.)\\M.M: E-mail: mona.mirzaei@nasa.gov\\  E.M.B.: E-mail: emily.m.barrentine@nasa.gov}

\begin{document} 
\maketitle

\begin{abstract}
The EXperiment for Cryogenic Large-aperture Intensity Mapping (EXCLAIM) is a cryogenic balloon-borne instrument that will map carbon monoxide and singly-ionized carbon emission lines across redshifts from 0 to 3.5, using an intensity mapping approach. EXCLAIM will broaden our understanding of these elemental and molecular gases, and the role they play in star formation processes across cosmic time scales. The focal plane of EXCLAIM’s cryogenic telescope features six \textmugreek-Spec spectrometers. \textmugreek-Spec is a compact, integrated grating-analog spectrometer, which uses meandered superconducting niobium microstrip transmission lines on a single-crystal silicon dielectric to synthesize the grating. It features superconducting aluminum microwave kinetic inductance detectors (MKIDs), also in a microstrip architecture. The spectrometers for EXCLAIM couple to the telescope optics via a hybrid planar antenna coupled to a silicon lenslet. The spectrometers operate from $420$--$540$~GHz with a resolving power $R = \lambda / \Delta \lambda = 512$, and employ an array of 355 MKIDs on each spectrometer. The spectrometer design targets a noise equivalent power (NEP) of $2\times10^{-18}$\,W/$\sqrt{\rm Hz}$ (defined at the input to the main lobe of the spectrometer lenslet beam, within a $9^\circ$ half width), enabled by the cryogenic telescope environment, the sensitive MKID detectors, and the low dielectric loss of single-crystal silicon. We report on these spectrometers under development for EXCLAIM, providing an overview of the spectrometer and  component designs, the spectrometer fabrication process, fabrication developments since previous prototype demonstrations, and the current status of their development for the EXCLAIM mission.
\end{abstract}


\section{INTRODUCTION}
\label{sec:intro}  

\begin{figure} [ht]
\begin{center}
\includegraphics[width=17cm]{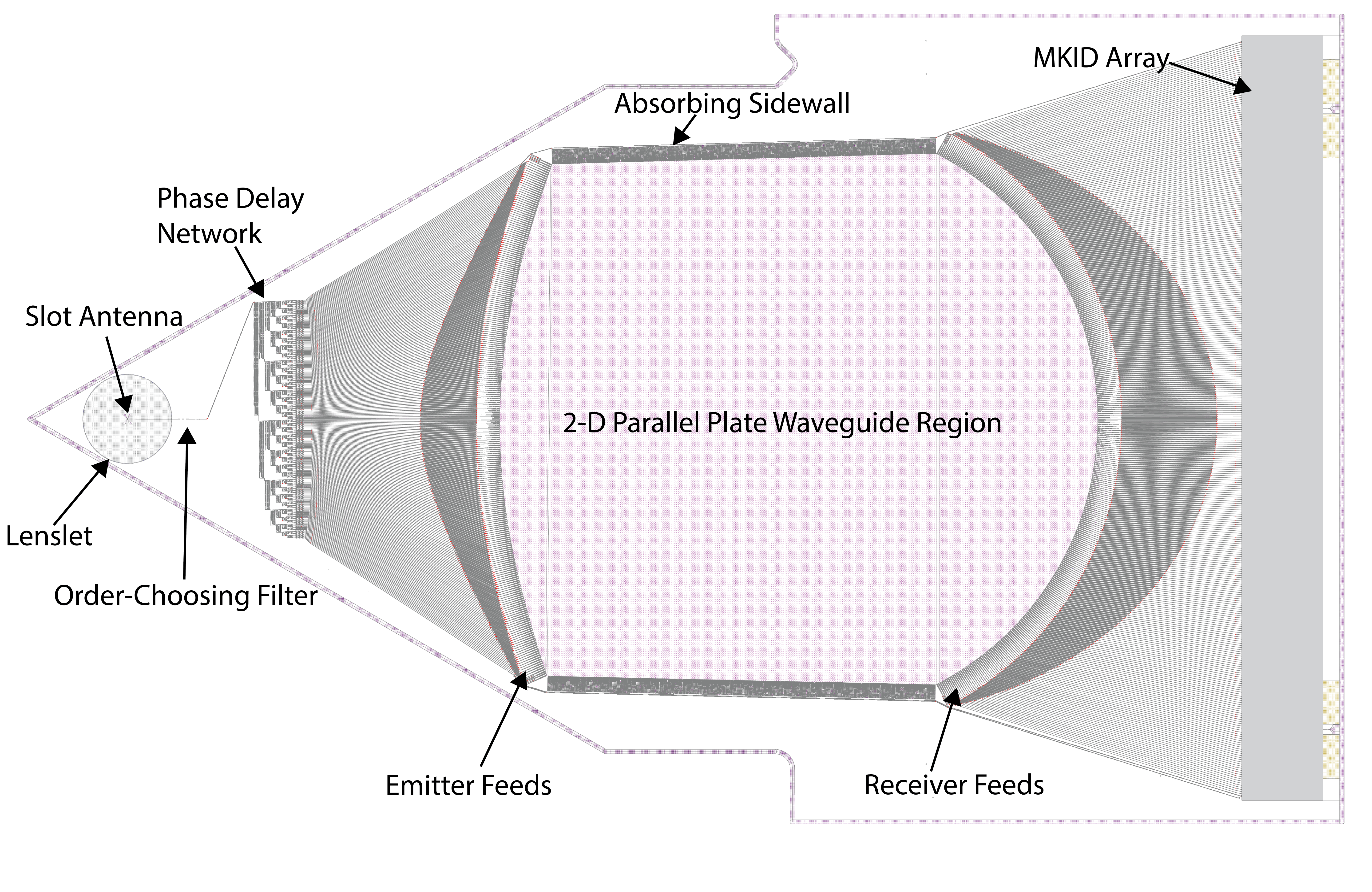}
   \end{center}
   \caption[] 
   { \label{fig:Spectrometer} 
The EXCLAIM \textmugreek-Spec spectrometer design. The spectrometer chip is 34 mm $\times$ 59 mm in size.}
 \end{figure} 

Multiwavelength imaging observations have revealed a wide range of environments in the universe \cite{Madau2014}. In the past 40 years, utilizing far-infrared and submillimeter spectroscopy, galaxies at redshifts as large as $z = 7.5$ have been confirmed with space and ground-based facilities \cite{Finkelstein2013}. These, and related optical observations, indicate an increase in the star formation rate from the cosmological reionization period up until redshifts of $z = 2$ \cite{Carilli2013,Madau2014,walter2020evolution}. After reaching this peak, the star formation rate falls by a factor of 10. 
To obtain a better understanding of the evolution of the physical conditions of galaxies across cosmic time, and how these influence the star formation rate, new measurements of the molecular and fine-structure emission lines of abundant elements such as carbon, nitrogen and oxygen are needed.
 
The EXperiment for Cryogenic Large-Aperture Intensity Mapping (EXCLAIM)\cite{ade2020experiment, EXCLAIMSPIE} is a new high-altitude balloon spectrometer mission that is designed to detect the submillimeter emission of redshifted carbon monoxide (CO) and singly-ionized carbon [CII] lines in windows spanning $0 < z < 3.5$.
Instead of resolving individual galaxies and measuring their brightness, EXCLAIM will measure the statistics of brightness fluctuations of redshifted, cumulative line emission, in a method known as intensity mapping (IM). It will provide a blind and complete survey of the emitting gas in a cross-correlation with the Baryon Oscillation Spectroscopic Survey (BOSS) \cite{alam2015eleventh}. IM reduces aperture and integration requirements relative to surveys that catalog individual galaxies, allowing the science goals to be achieved in a conventional balloon flight.
The optimal band for BOSS cross correlation is $420$--$600$~GHz; however, a bright ortho-water emission line in the upper atmosphere occurs at $557$~GHz. To avoid that emission\cite{Carilli2013}, EXCLAIM operates from $420$--$540$~GHz. 

EXCLAIM fields six \textmugreek-Spec spectrometers on an all-cryogenic (1.7~K) telescope. \textmugreek-Spec \cite{barrentine2016design,cataldo2015four,cataldo2014micro,noroozian2015mu} integrates all the elements of a grating spectrometer onto a silicon chip. Due to the high index of silicon, this provides an order of magnitude reduction in size compared to grating spectrometer designs that employ a free-space grating~\cite{Yen,Bradford2004}.It should be noted that \textmugreek-Spec is similar, but differs due to its grating-analog design, to other on-chip superconducting spectrometers under development, which instead use a filter-bank design~\cite{Shirokoff, Endo}. We report here on the status of the \textmugreek-Spec spectrometers under development for EXCLAIM, including the EXCLAIM spectrometer and component designs, expected performance, fabrication development and the planned spectrometer fabrication process. 

\section{The EXCLAIM \textmugreek-Spec Design }

 The EXCLAIM \textmugreek-Spec design operates over $420$--$540$~GHz, with a resolving power of $R=\lambda/ \Delta \lambda \sim512$ and an expected noise equivalent power (NEP) of $2\times10^{-18}$\,W/$\sqrt{\rm Hz}$, defined at the input to the main lobe of the spectrometer lenslet beam (within a $9^\circ$ half width), when under the typical expected sky loading in the primary science channels ($0.7$ fW at the spectrometer lenslet input, assuming an optical efficiency of $74\%$ through the telescope optics). A schematic of the EXCLAIM \textmugreek-Spec design is shown in Figure~\ref{fig:Spectrometer}, and a summary of the EXCLAIM \textmugreek-Spec design and performance parameters is shown in Table~\ref{Spectdesigntable}. The integrated \textmugreek-Spec spectrometer design is realized in niobium planar transmission line on a thin (450~nm thick) and low-loss\cite{datta2013large,loewenstein1973optical,afsar1994millimeter, wollack2020infrared} single-crystal silicon dielectric, and is based on the \textmugreek-Spec design approach previously demonstrated as an $R=64$ prototype\cite{cataldo2014micro,noroozian2015mu}.  

 Light is coupled to the spectrometer from the optics of the EXCLAIM telescope via an on-chip planar slot antenna with a silicon lenslet. The slot antenna design is based on a cross-slot design approach\cite{iacono2011line}, and operates over a full octave ($300$--$600$~GHz). The slot antenna is located at the focus of a 4-mm-diameter hyper-hemispherical silicon lenslet, which is attached to the back of the spectrometer silicon chip. A parylene-C anti-reflection (AR) coating is applied to the spherical surface of the silicon lenslet, with a thickness optimized to provide optimal coupling over the EXCLAIM band. Further details of the EXCLAIM optics design are provided in Ref.~\citenum{EXCLAIMopticsSPIE} in these proceedings.
 
\begin{table}
\caption{The EXCLAIM \textmugreek-Spec spectrometer and MKID design parameters.} 
\label{Spectdesigntable}
\begin{center}   

\begin{tabular}{|m{8.5cm}|m{7.5cm}|}
\hline
  Number of spectrometers
 & 6 \\
\hline
  Spectrometer spectral band
 & $420$--$540$~GHz 
 \\
\hline
Spectrometer grating order, $M$ & 2 (single order) \\
\hline
Spectrometer resolving power, $R$ 
 & $512$ at 472~GHz (center frequency) \\  & $535$--$505$ over spectral band  \\
\hline
Spectrometer efficiency  & $23\%$ \\
\hline
Spectrometer NEP (at input to lenslet main lobe) & $2\times10^{-18}$\,W/$\sqrt{\rm Hz}$ at 0.70~fW (typical loading) \\
\hline
MKID NEP (at input to each MKID)  & $5\times10^{-19}$\,W/$\sqrt{\rm Hz}$ at 0.16~fW (typical loading)\\
\hline
Number of receivers/MKIDs per spectrometer & 355 \\
\hline
MKID readout band & $3.25$--$3.75$~GHz \\
\hline
Operating temperature & $100$~mK \\
\hline
\end{tabular}
\end{center}
\end{table}

 An impedance-transforming microstripline feed couples to the slot antenna and transmits the signal to the microstripline phase delay network. Here the signal is split in $N=256$ branches, in a binary tree configuration using broadband ($300$--$600$~GHz) 4-stage Wilkinson power dividers. A linear phase delay gradient is applied by adjusting the length of meandered microstripline in this power splitting network. The resulting phase-delayed wavefronts are then launched via emitter feeds into a 2D parallel-plate waveguide, and recombine constructively at receiver feeds located at angles along the spectrometer focal plane as a function of wavelength in a Rowland circle configuration~\cite{rotman1963wide,rowland1883xxix} with a Rowland circle radius of 1.35 cm. These receiver feeds Nyquist sample the spectral function of the synthesized grating, which is well approximated by a sinc$^{2}$ function. 
 The focal plane receiver and emitter locations and the transmission line lengths of the phase delay network are optimized~\cite{microspec3} taking into account dispersion effects due to the kinetic inductance of the niobium planar transmission line and due to the geometry and bends of the meandered delay network. The length of the microstrip line between the power splitters in the binary tree network is also designed to avoid the occurrence of stopbands in the EXCLAIM band due to coherent reflections off of the bends~\cite{barrentine2016design}. In addition, the angles of the receivers are constrained to viewing angles of {$\leq 45\%$} with respect to the emitter array to optimize coupling. This focal plane and delay network design spans a single grating order, $M=2$, provides diffraction-limited phase error over the spectrometer band and spectral resolving power of $R = NM = 512$ at the center frequency of the EXCLAIM spectral band. The order of the spectrometer grating is selected by an on-chip stepped-impedance microstripline ``order-choosing" bandpass filter with a passband of $345$--$603$~GHz located at the input of the spectrometer phase delay network, by low- and high-pass metal-mesh filters inserted in the EXCLAIM telescope optics, and by the niobium transmission lines, which terminate light above the superconducting energy gap at ${\sim}680$~GHz.  
 
  \begin{figure} [ht]
\begin{center}
\includegraphics[width=17cm]{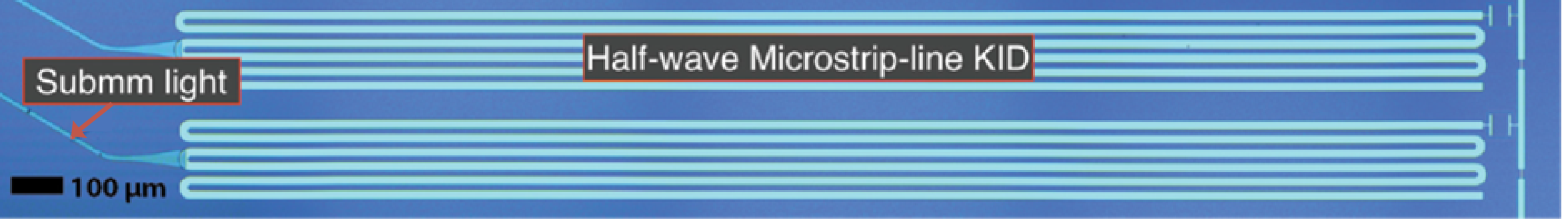}
   \end{center}
   \caption[] 
   { \label{fig:MKIDapproach} 
An aluminum half-wave microstrip transmission line design will be employed for the MKIDs in the EXCLAIM \textmugreek-Spec design, similar to the design used for the $R=64$ \textmugreek-Spec prototype MKID array, which is shown above.}
   \end{figure} 
 
Each receiver feed transitions back to microstrip and to an individual MKID detector. There are 355 receivers and MKIDs on each spectrometer and all of the MKIDs on a spectrometer are read out on a single 50-Ohm aluminum microstrip microwave feedline, with the MKIDs coupled via a parallel-plate coupling capacitor formed between the top aluminum layer and an isolated region of the niobium ground plane layer. The MKIDs consist of two branches of a half-wave microstrip transmission-line resonator, with a 20-nm-thick aluminum microstrip absorbing layer and a niobium ground plane. Their resonance frequencies span $3.25$--$3.75$~GHz. The sub-millimeter optical input lies at the microwave zero voltage node, between two branches of symmetric half-wave transmission line resonators, providing isolation between the microwave and sub-millimeter optical signals. This EXCLAIM MKID design is similar to the one shown in Figure~\ref{fig:MKIDapproach}, which was used in the $R=64$ \textmugreek-Spec demonstration, but features a thinner and narrower aluminum microstripline for increased sensitivity. The spectrometer and MKIDs operate at a 100-mK bath temperature. Aluminum MKIDs have demonstrated excellent sensitivity with NEPs as low as $\sim 10^{-19}$~W/$\sqrt{\rm Hz}$ \cite{baselmans2017kilo}. Based upon our measurements of microwave loss, two-level system (TLS) noise and quasiparticle heating effects in representative $\sim 20$-nm-thick aluminum coplanar waveguide (CPW) resonator structures (similar to the niobium CPW resonator structures described in Section~\ref{fabdevelop}), our design model predicts an MKID NEP of $\sim 5 \times 10^{-19} $~W/$\sqrt{\rm Hz}$, under the typical expected loading in the primary science channels ($0.16$~fW at the input to each MKID). 

The expected efficiency of the spectrometer from the input of the lenslet to the input of each MKID is estimated to be $\sim23\%$. Sources of loss include the optical coupling of the slot-antenna lenslet beam to the EXCLAIM receiver optics (estimated loss of $51\%$), loss in the coupling between the receiver and emitter feeds in the spectrometer focal plane (estimated loss of $50\%$, due to limitations of the chosen feed design), and loss in the planar transmission lines dominated by loss in the silicon dielectric (estimated loss of $\leq5\%$). It should be noted that this estimate assumes loss through the spectrometer planar transmission lines arising from the superconductor and the superconductor interfaces themselves is minimal. As will be described in Sec.~\ref{fabdevelop}, in previous \textmugreek-Spec prototype devices, this was not the case, and because of this we implement changes to the spectrometer fabrication process. Ultimately, planned measurements of sub-millimeter loss in the first fabricated \textmugreek-Spec devices for EXCLAIM will provide a better understanding of this loss term.

The geometry of the microstrip transmission line provides high immunity to stray light and cross-talk, due to the thin 450-nm-thick silicon dielectric layer and the protective niobium ground plane. In the 2D parallel plate waveguide cavity, a meta-material absorbing sidewall structure\cite{bulcha2018electromagnetic} provides $\ge 99\%$ attenuation of signals over a full octave ($300$--$600$~GHz) and over a range of incidence angles (45-90 degrees). This sidewall absorber structure provides additional termination for any out-of-band signals and of any reflections in the 2D waveguide cavity. In addition, a thin film titanium absorber with sheet resistance $\sim150$ Ohms/square is deposited on the back of the spectrometer silicon chip to limit the coupling of stray power to the detectors through the silicon. Thermal blocking filters~\cite{wollack2014impedance} can be added at the input to the microwave readout lines of the spectrometer package. The spectrometer focal plane package design also features a baffle structure to reduce the field of view, and absorbing sidewalls, to attenuate stray light before it reaches the sensitive detectors.

\section{Spectrometer Fabrication}
\label{SpecFab}

The  EXCLAIM \textmugreek-Spec fabrication follows a process similar to that used for the $R=64$ prototypes devices, though new developments to the process will be discussed in Section~\ref{fabdevelop}. In this process, superconducting aluminum and niobium films are patterned on both sides of the single-crystal silicon device layer of a silicon-on-insulator (SOI) wafer, using a low-temperature flip-bonding process\cite{patel2013fabrication}. The process begins with an SOI wafer with a float-zone (f-z) silicon device layer of 450-nm thickness for the dielectric. The niobium ground plane layer is fabricated via a lift-off process (similar to that described in Refs.~\citenum{patel2013fabrication,brown2018fabrication}) to avoid roughening and etching through the silicon device layer. The niobium lift-off process is performed using germanium as a hard mask. This process starts by applying hexamethyldisilazane (HMDS) and a thin layer of photoresist ($0.5$--$1$\textmu m) followed by a soft bake to remove the solvent. A thin layer of germanium is then deposited via evaporation or sputtering. This germanium layer is masked by another layer of HMDS and photoresist, which are baked at low temperature over a long duration and are exposed via a contact mask. The lower temperature of the bake protects the germanium from thermal shock and the long time is needed to remove all the solvents in the resist. Then the germanium is etched with Reactive Ion Etching (RIE). The photoresist is then ashed in an O\textsubscript{2} plasma to remove the top layer and define the undercut in the lift-off profile. The niobium ground plane is deposited via sputtering, while the silicon native oxide is etched just prior to deposition. In the final step, the niobium pattern is lifted-off by acetone. 

A benzocyclobutene (BCB) polymer resin is used to bond the niobium ground plane side of the SOI wafer to the front side of the f-z silicon wafer. The SOI handle wafer is removed by a combination of lapping and etching. The buried oxide layer is removed with a Buffered Oxide Etch (BOE) to expose the other side of the silicon device layer. A similar lift-off process with a germanium hard mask is then used to define the top niobium microstrip circuitry while preventing damage to the silicon dielectric layer. For this top-side niobium patterning, a two-step lithography process using a contact mask and direct writing, is performed. The direct write is used to obtain the sub-micron feature sizes required for the slot antenna feed design. A gold-palladium (AuPd) layer (sheet resistance of $\sim20$ Ohms/square) with a titanium (Ti) adhesion layer is then deposited between, and over, regions of the niobium microstrip lines using a resist lift-off mask. This Ti/AuPd layer forms the absorber in the spectrometer focal plane sidewalls, and in the power dividers in the phase delay network. The next step in the process is to pattern the aluminum layer which forms the MKID array structure. To ensure the proper aluminum film profile and a clean silicon interface, the native oxide on the silicon is removed by a hydrogen fluoride (HF) dip and a reverse bias clean is completed just prior to the aluminum deposition (see further details of this process in Section~\ref{fabdevelop}). After sputter-depositing the aluminum, HMDS and a thin resist layer are applied for the lithography. The photoresist is patterned via either a contact mask or a combination of contact mask and direct writing, followed by a gentle develop step. Protected by the resist, the aluminum is wet etched in aluminum etchant and the remaining resists are stripped in solvent. 

\begin{figure}[htp]
\begin{center}
\centering
\begin{subfigure}[b]{0.22\textwidth} \centering 
\includegraphics[width = \textwidth]{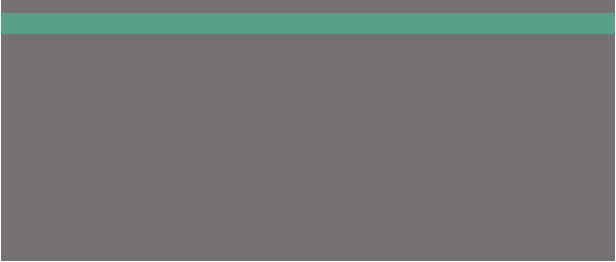} \captionsetup{width = \textwidth}
\caption{} \end{subfigure}
\centering
\begin{subfigure}[b]{0.22\textwidth} \centering
\includegraphics[width = \textwidth]{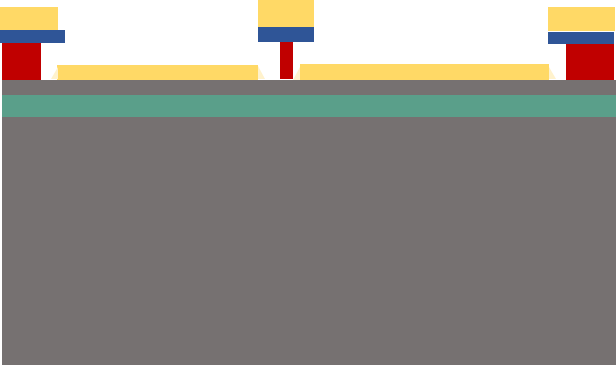}
\captionsetup{width = \textwidth} \caption{}
\end{subfigure}
\centering
\begin{subfigure}[b]{0.22\textwidth}  \centering
\includegraphics[width = \textwidth]{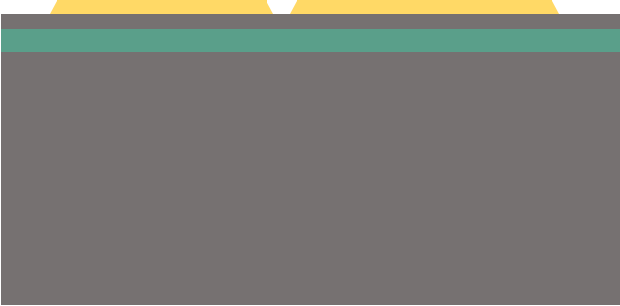} \captionsetup{width = \textwidth}
\caption{}
\end{subfigure}
\centering
\begin{subfigure}[b]{0.22\textwidth} \centering
\includegraphics[width = \textwidth]{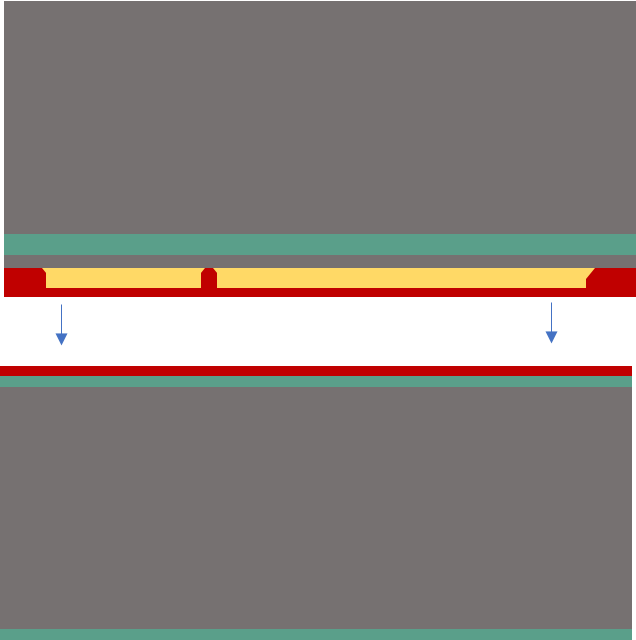} \captionsetup{width = \textwidth}\caption{} \end{subfigure}
\centering
\\
\begin{subfigure}[b]{0.22\textwidth}  \centering
\includegraphics[width = \textwidth]{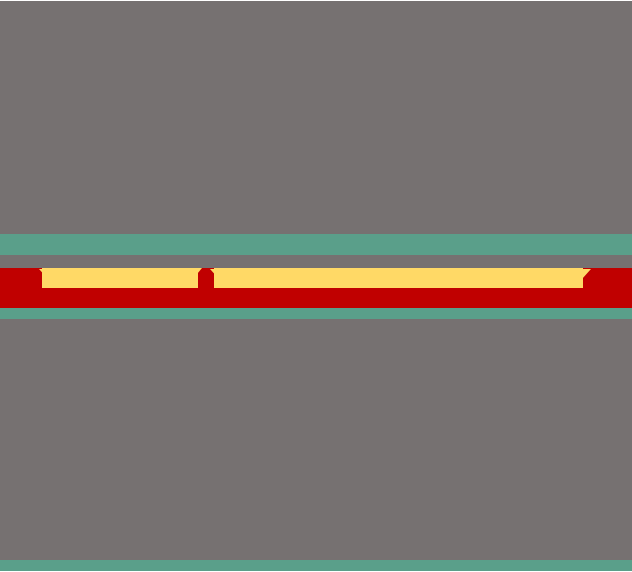}
\captionsetup{width = \textwidth} \caption{} \end{subfigure}
\centering
\begin{subfigure}[b]{0.22\textwidth}
\centering
\includegraphics[width = \textwidth]{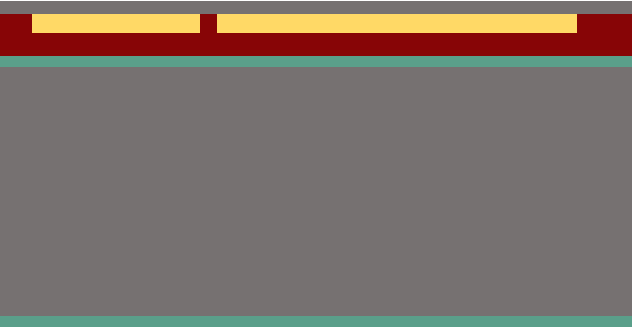}
\captionsetup{width = \textwidth} \caption{} \end{subfigure}
\centering
\begin{subfigure}[b]{0.22\textwidth} \centering 
\includegraphics[width = \textwidth]{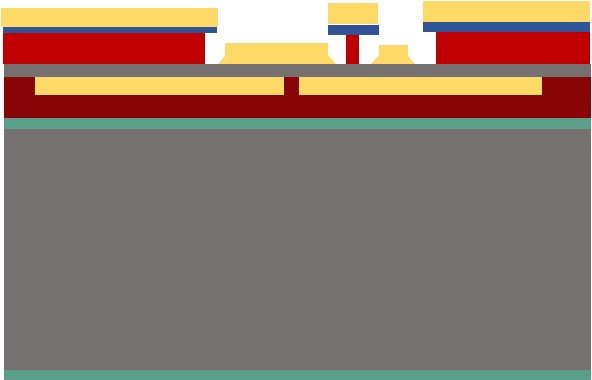} \captionsetup{width = \textwidth} \caption{} \end{subfigure}
\centering \begin{subfigure}[b]{0.22\textwidth}  
\centering 
\includegraphics[width = \textwidth]{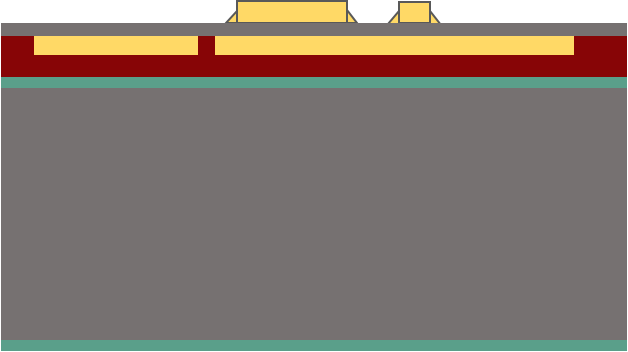}
\captionsetup{width = \textwidth} \caption{} \end{subfigure}
\\
\begin{subfigure}[b]{0.22\textwidth} \centering 
\includegraphics[width = \textwidth]{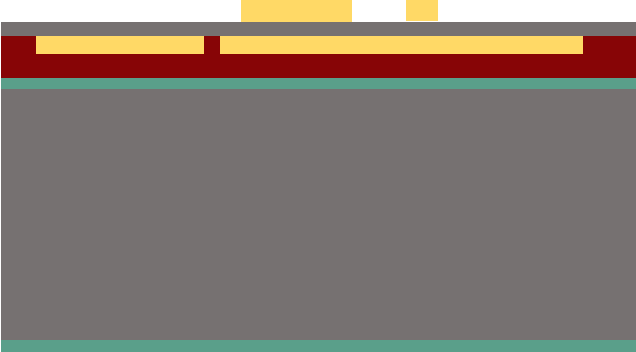} \captionsetup{width = \textwidth}
\caption{} \end{subfigure}
\centering
\begin{subfigure}[b]{0.22\textwidth} \centering
\includegraphics[width = \textwidth]{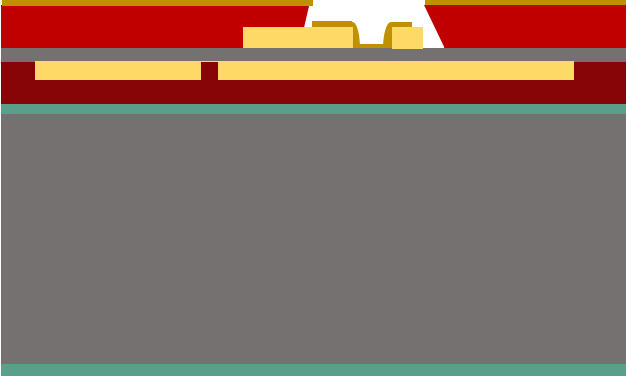}
\captionsetup{width = \textwidth} \caption{} 
\end{subfigure}
\centering
\begin{subfigure}[b]{0.22\textwidth} \centering
\includegraphics[width = \textwidth]{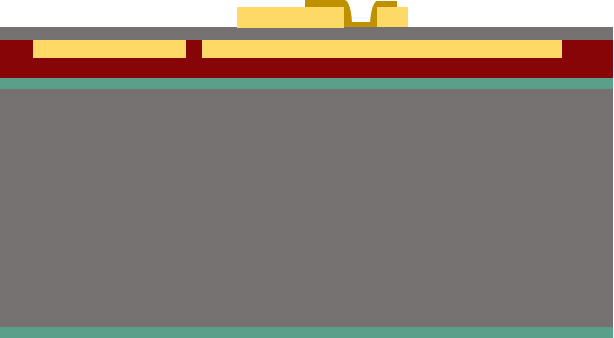} \captionsetup{width = \textwidth}\caption{} \end{subfigure}
\centering
\begin{subfigure}[b]{0.22\textwidth}  \centering
\includegraphics[width = \textwidth]{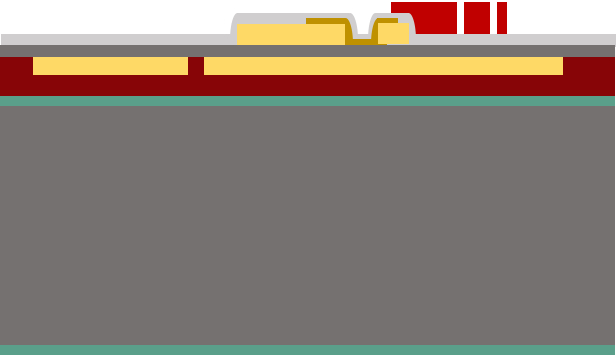}
\captionsetup{width = \textwidth} \caption{} \end{subfigure}
\\
\centering
\begin{subfigure}[b]{0.22\textwidth} \centering
\includegraphics[width = \textwidth]{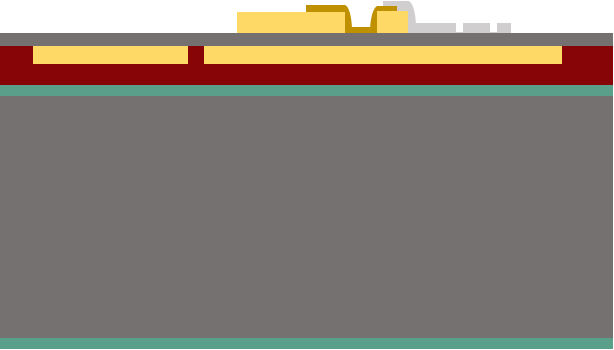}
\captionsetup{width = \textwidth} \caption{} \end{subfigure}
\centering
\begin{subfigure}[b]{0.22\textwidth}  \centering \includegraphics[width = \textwidth]{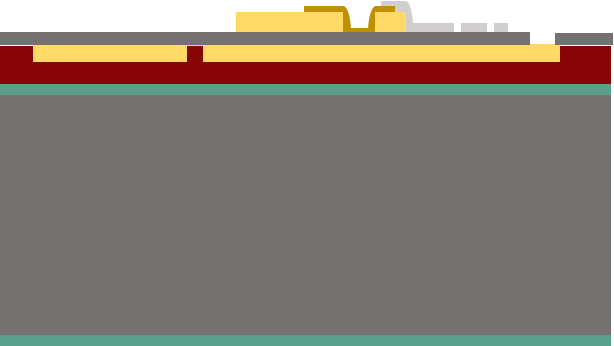} \captionsetup{width = \textwidth} \caption{} \end{subfigure}
\centering
\begin{subfigure}[b]{0.22\textwidth} \centering \includegraphics[width = \textwidth]{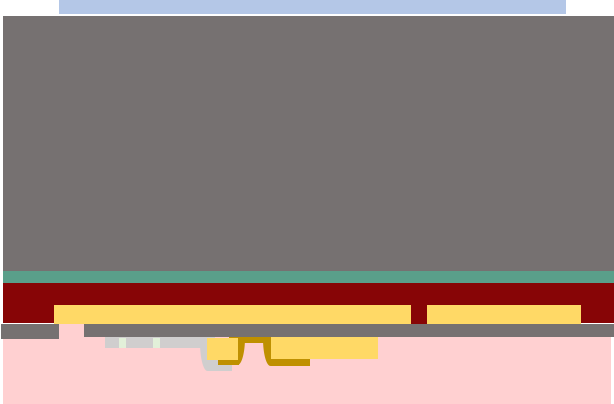} \captionsetup{width = \textwidth} \caption{} \end{subfigure} \centering
\begin{subfigure}[b]{0.22\textwidth}  \centering 
\includegraphics[width = \textwidth]{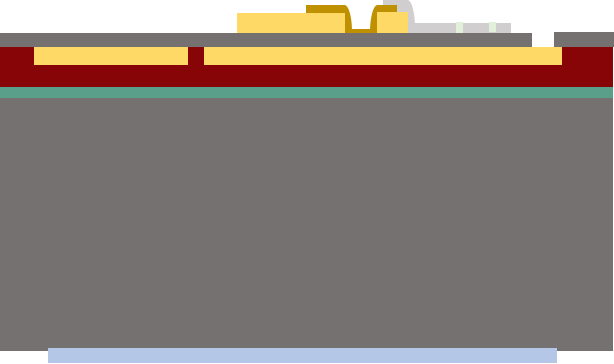} \captionsetup{width = \textwidth} \caption{} \end{subfigure}
\\
  \centering
\begin{subfigure}[b]{0.7\textwidth}          \centering 
\includegraphics[width = \textwidth]{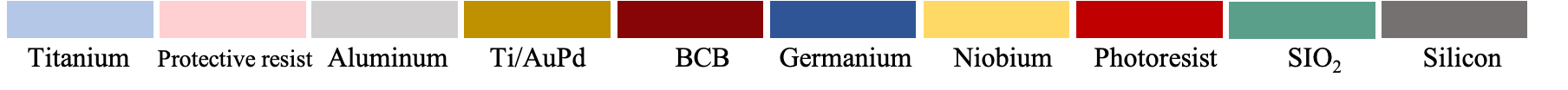} \captionsetup{width = \textwidth}  \end{subfigure}
    \centering
\captionsetup{width = 17cm} \caption{The planned spectrometer fabrication process flow. a) Begin with an SOI wafer. b) A photo-resist/germanium hard mask is patterned. The germanium is etched by RIE. Then masked by the germanium, the resist underlayer is ashed in an O\textsubscript{2} plasma to remove the resist layer and define the mask undercut. The niobium ground layer is deposited via sputtering. The silicon (Si) native oxide is etched just prior to deposition. c) The lift-off process is completed to define the niobium ground plane layer. d) and e ) The niobium ground plane side of the wafer is bonded to a fz-silicon backing wafer coated with thermal oxide using a BCB polymer resin. f) The SOI handle wafer is removed by a combination of lapping and etching. The SOI buried oxide layer is removed with BOE. g) A photo-resist/germanium hard mask for the top-side niobium circuitry is patterned. The germanium hard mask is etched by RIE. Masked by the germanium, the resist underlayer is ashed in an O\textsubscript{2} plasma to remove the top layer and define the undercut. The top niobium layer is sputtered-deposited through the mask, with removal of the native oxide layer just prior to deposition. h) The top niobium lift-off is completed. Then a self-limited layer of niobium oxide is grown on the surface and subsequently removed by wet etching. i) This oxidation-wet etch cycle is repeated as necessary to remove the extended sidewalls.  j) A lift-off resist mask is patterned and the Ti/AuPd absorber layer is deposited. k) The Ti/AuPd layer is lifted off. l) After sputter-depositing the aluminum, the photoresist is applied and patterned. m) The aluminum layer is wet etched. n) The SOI silicon device layer is etched through for access to the ground plane. o) The wafer frontside is protected by resist and the backside thermal oxide on the silicon backing wafer is removed, followed by patterning of the titanium absorber layer. p) The spectrometer wafer is released in acetone.}
\label{process flow}
\end{center}
\end{figure}

In the next step, the device layer is etched to provide access to the ground plane niobium for wirebonding purposes. Then a protective resist is applied to protect the wafer frontside and the titanium resistive layer is deposited on the back side of the supporting silicon backing wafer. The titanium is etched back by BOE, while it is masked by the photoresist. In the last step, the final release of the wafer is performed in acetone. This full process is illustrated in Figure~\ref{process flow}. 

\section{Fabrication Development}
\label{fabdevelop}

\begin{figure}[htp]
  \centering
  \subfloat[]{\includegraphics[width=0.47\textwidth]{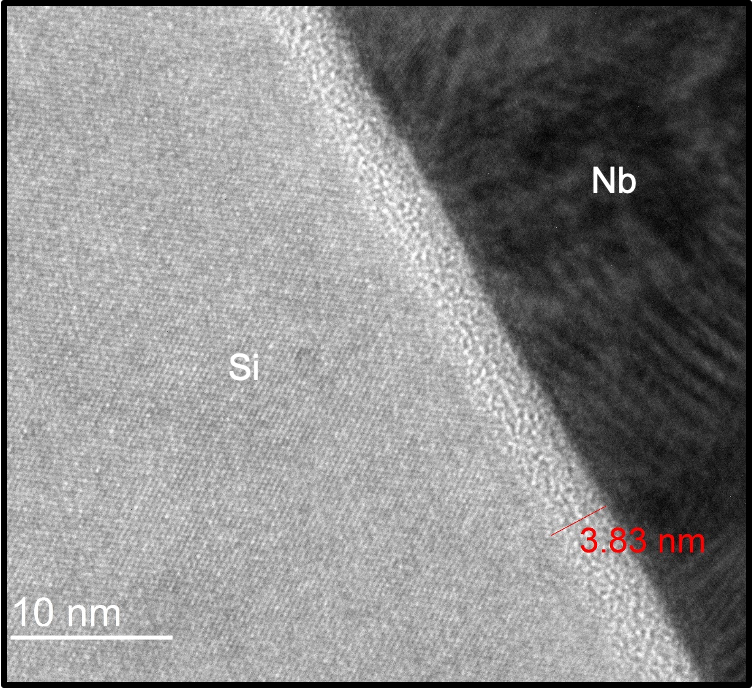}\label{fig:amorphous}}
  \hfill
  \subfloat[]{\includegraphics[width=0.43\textwidth]{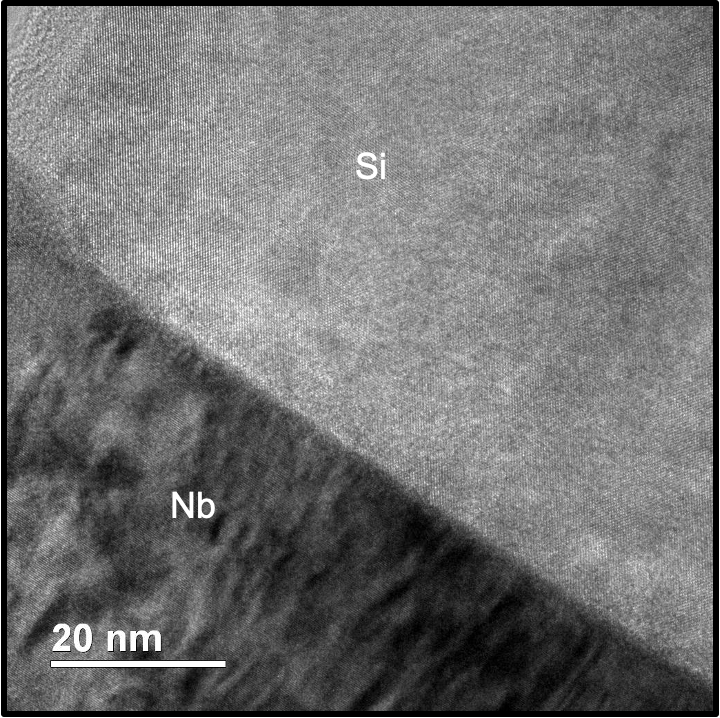}\label{fig:clean TEM}}
 
  \caption{a) {Transmission electron microscopy (TEM)} image showing the ground plane niobium (Nb)-silicon (Si) interface of an $R=64$ µ-Spec prototype. Here
there is evidence of an amorphous layer which was a source of TLS loss.
b) TEM image showing the Nb-Si interface of a
CPW resonator, which
underwent a modified niobium lift-off process with a BOE clean of the native oxide prior to the niobium deposition and removal of the extended sidewalls. Here the contact is sharp, with no evidence of amorphous layers.}
\end{figure}

\begin{figure}[htp]
  \centering
  \subfloat[]{\includegraphics[width=0.47\textwidth]{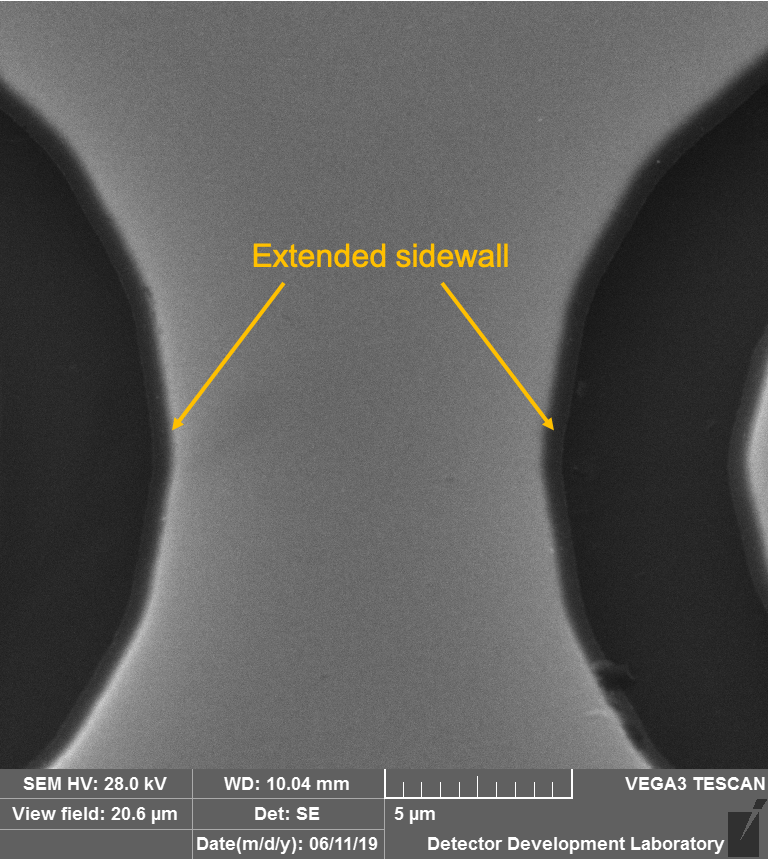}\label{fig:Nbtail1}}
  \hfill
  \subfloat[]{\includegraphics[width=0.47\textwidth]{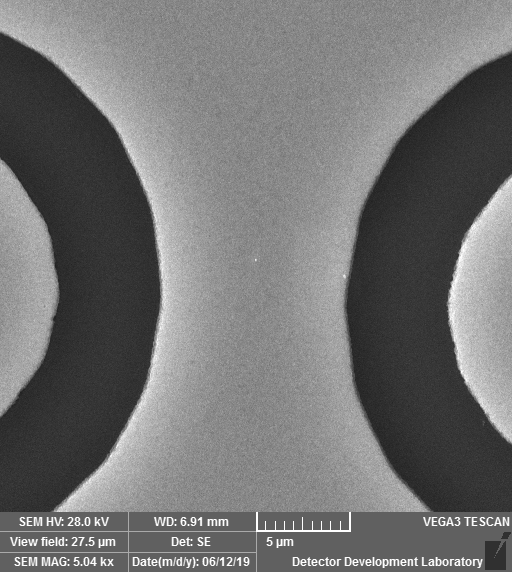}\label{fig:Nbtail2}}
 
  \caption{a) SEM image of the as deposited niobium lift-off pattern. Note the thin extended sidewalls at the edges of the pattern, which have
been observed to result in higher
microwave loss and low $Q_{i}$ in CPW resonator
structures. b) After successive oxidation and
etch cycles the extended sidewalls are
removed and the resulting
measured $Q_{i}$ is more than 10$\times$ higher (see Table~\ref{sample comparison}).} 
\end{figure}

A two-layer lift-off process was developed\cite{patel2013fabrication} for the $R=64$ \textmugreek-Spec prototype to pattern the thin film superconducting niobium microstrip transmission lines (as described in Section~\ref{SpecFab}) and resulted in a US patent\cite{brown2015high}. This original technique provided the precise control of linewidth ($\leq\pm0.2$~\textmu m) required to ensure phase control in the microstrip transmission lines and thus the high resolution of the spectrometer. However, results of the initial prototype $R=64$ \textmugreek-Spec spectrometers, and separate diagnostic CPW resonator devices, showed that unexpected loss was being introduced due to the lift-off process. This extra loss was determined to be due to thin extended sidewalls of the niobium, at the edge of the niobium traces, resulting from the sputter-deposited lift-off process (Figure~\ref{fig:Nbtail1}), and an amorphous native oxide layer at the niobium-silicon substrate interface (Figure~\ref{fig:amorphous}).  

\begin{figure}[htp]
\centering
  \includegraphics[scale=0.467]{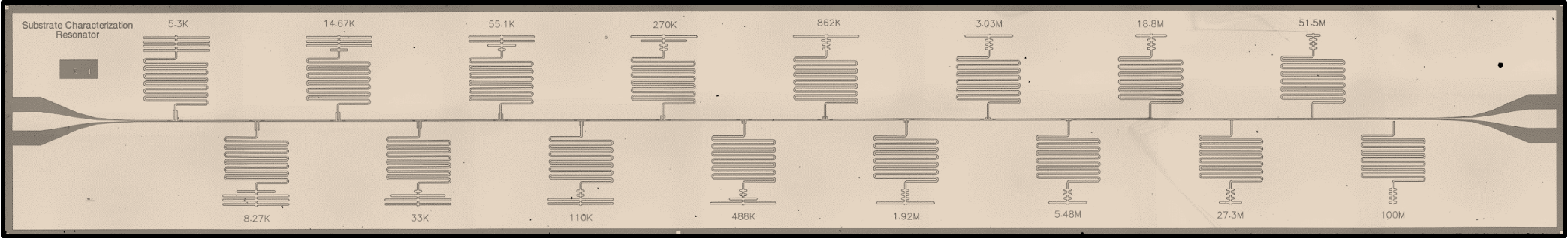}
  \caption{Image of a niobium CPW resonator chip, featuring quarter-wave resonators with ``fishbone" tuning structures. This resonator chip design was used as a diagnostic device, to measure microwave loss in response to variations in the niobium lift-off process.}
  \label{fig:msquid}
\end{figure}

An improved lift-off technique using a two-step selective etching method was developed to provide a clean metal-silicon interface and to remove the lossy extended sidewalls. In this modified lift-off process, the amorphous native oxide layer on the silicon substrate is removed by either a BOE or a reverse bias cleaning process, just prior to the deposition of the niobium film (Figure~\ref{fig:clean TEM}). After the lift-off is complete, a self-limited layer of niobium oxide is grown on the surface and subsequently removed by wet etching to remove the thin extended tails. The oxidation-wet etch step is repeated for as many cycles as necessary (Figure~\ref{fig:Nbtail2}). In developing this modified lift-off process, niobium CPW quarter-wave resonators with a ``fishbone" resonance frequency tuning structure\cite{stevenson2009superconducting} (Figure~\ref{fig:msquid}) were used for a rapid turn-around study of the impacts of process variations on internal microwave quality factors, $Q_{i}$. Coupling quality factor, $Q_{c}$, values for these diagnostic resonators ranged from ${\sim}5,000$ to ${\sim}500,000$ and resonance frequencies were at ${\sim}3.5$~GHz. The CPW readout feedline in these devices was formed from the same niobium layer and there were no subsequent fabrication steps. These CPW devices were tested at base temperatures of $7$--$30$~mK, inside a magnetically-shielded and light-tight dilution refrigerator. Table~\ref{sample comparison} reports the microwave loss of  four individual devices, with or without sidewall removal and pre-cleaning of the native oxide. For comparison purposes, measurements for etched niobium resonator samples, previously reported in Ref.~\citenum{brown2016impact}, are also shown. The modified lift-off process results in a decrease in microwave loss by more than an order of magnitude when measured in these CPW microwave resonator structures. 

  \begin{table}
\caption{Comparison table of microwave loss and residual resistivity ratio (RRR) measurements for niobium CPW resonators chips with modifications to the niobium lift-off process. Also included for comparison are etched niobium devices. All devices measured were CPW resonators with a fishbone tuning structures (see Figure~\ref{fig:msquid}). } 
\label{sample comparison}
\begin{center}   
\begin{tabular}{|m{5cm}|m{4.85cm}|m{4.7cm}|m{0.85cm}|}

\hline
  \textbf{Primary Process Differences} & \textbf{Microwave $Q_{i}$ \newline Readout power $=-130$ dBm} &\textbf{Microwave $Q_{i}$ \newline Readout power $=-80$ dBm} & \textbf{Film  RRR} \\
\hline
  * Reactive ion etch of Nb &  400,000-1,000,000 & 200,000-400,000 & 5.9  \\
\hline
  * Nb lift-off \newline * BOE clean prior to deposition  \newline
  * Removal of sidewalls  &
200,000 & 100,000-150,000 & 5.7 \\
\hline
  * Nb lift-off \newline
  * Reverse bias prior to deposition \newline
  * Removal of sidewalls  &
150,000-350,000 & 100,000-200,000 & 5.0 \\
\hline
 * Nb lift-off  \newline
 * BOE clean prior to deposition \newline
 * No removal of sidewalls  &
10,000-20,000 & 7,000-10,000 & 7.0 \\
\hline 
* Nb lift-off \newline
* No native oxide removal  \newline
* No removal of sidewalls
 &
4,000-8,000 & 4,000-8,000 & 6.3 \\
\hline

\end{tabular}
\end{center}
\end{table}

\begin{figure}[htp]
  \centering
  \subfloat[]{\includegraphics[width=0.47\textwidth]{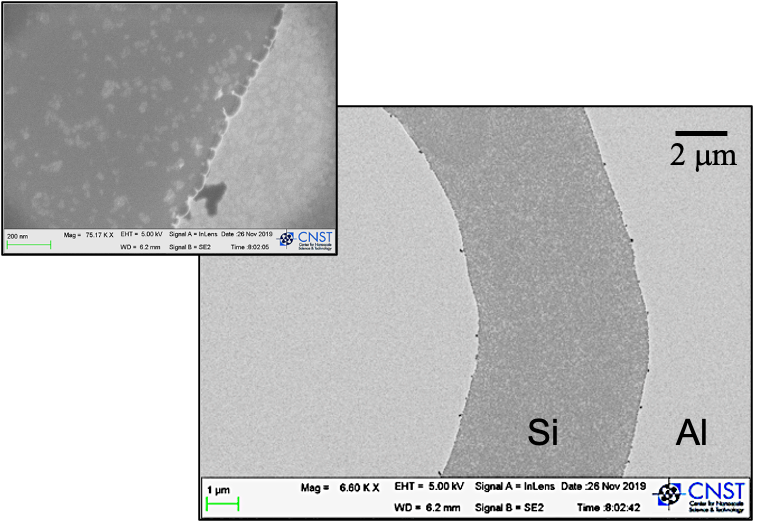}\label{fig:Alnetch}}
  \hfill
  \subfloat[]{\includegraphics[width=0.47\textwidth]{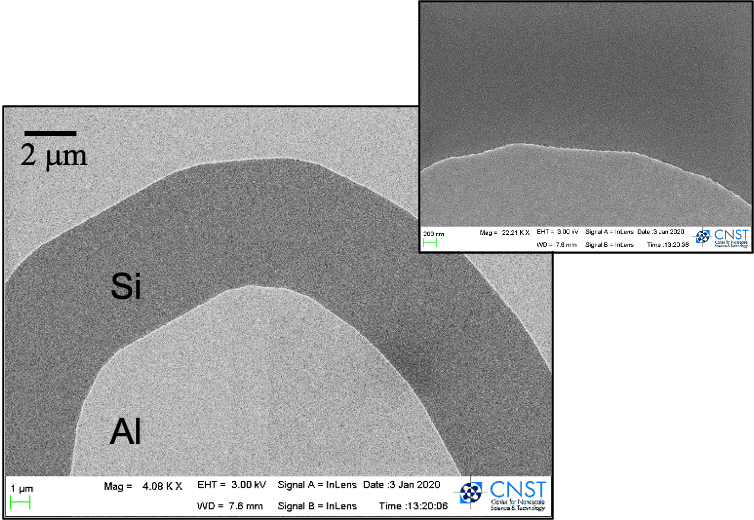}\label{fig:Aletch}}
 
  \caption{a) SEM image of a wet-etched aluminum layer on silicon. Note the residual aluminum in the etched regions in the absence of cleaning prior to the deposition. b) Applying a reverse bias prior to the aluminum deposition improves the quality of the etch and clears the residual aluminum in the etched area.}
\label{fig:Alpattern}
\end{figure}

The EXCLAIM MKID arrays employ a $20$\,nm thick aluminum layer, with narrower linewidths than were required for the aluminum MKID arrays in the $R=64$ prototype. In the design of the microstrip transmission line for the resonators, the coupling capacitor structures and the microstrip feedline, aluminum linewidths of $\sim$0.8 to 1.7 \textmu m were originally considered, though the design has since converged to minimum linewidths of $\sim$1.3~\textmu m. To fabricate these features with precise control on the width variation and to protect the thin single-crystal silicon dielectric from roughening and etching, contact and high-resolution direct laser writing lithography patterning steps, followed by a wet aluminum etch process, have been developed. These processes need to ensure contact between the spectrometer sub-millimeter niobium transmission lines and the aluminum MKID structure. Proper step coverage and surface cleaning is important to obtain a low resistance electrical contact and low microwave loss. To control the etch rate of the aluminum for sub-micron-to-micron feature sizes, the aluminum etch is performed at room temperature, which reduces the etch rate to 40 \angstrom/sec compared to the nominal etch rate of 125 \angstrom/sec at $50^\circ$C. Figure~\ref{fig:Alpattern} shows scanning electron microscope (SEM) images of the silicon surface after the aluminum wet etch process. Initially, evidence of residual aluminum (which could lead to TLS noise and loss) was seen after the completion of the etch. Additional substrate cleaning with a reverse bias prior to the aluminum deposition was found to be a critical step to address this issue. 

\section{Current Status}
{Fabrication of the EXCLAIM \textmugreek-Spec spectrometers began in October 2020 and will follow the processes described in Sections~\ref{SpecFab} and~\ref{fabdevelop}. Characterization of the resulting individual spectrometers and their MKID arrays, including measurement of the NEP, resolving power, and absolute frequency response of the spectrometer channels is planned prior to their integration into the EXCLAIM receiver and instrument. The first EXCLAIM balloon flight is anticipated in late summer or fall 2022.}

\acknowledgments 
This work was supported by funding provided by the NASA Astrophysics Research and Analysis (APRA) Program.  We gratefully acknowledge the previous leadership and guidance of S. Harvey Moseley in the development of the high-resolution \textmugreek-Spec technology. We also acknowledge the previous contributions of Kongpop U-Yen towards the development of a higher-resolution \textmugreek-Spec design, including a new design of the broadband power divider, which is employed in the high-resolution EXCLAIM spectrometer design. We also acknowledge the founding contributions of Ari Brown in the development of the $R=64$ prototype spectrometer fabrication process, which forms the basis for the EXCLAIM spectrometer fabrication process described here. We thank Ken Livi of the Materials Science and Engineering Department at Johns Hopkins University for completing TEM imaging and interpretation of the niobium samples described in Sec.~\ref{fabdevelop}. We also thank Sang Yoon and Aveek Gangopadhyay for assistance with the fabrication of aluminum and niobium resonators samples. Fabrication research was performed in part at the NIST Center for Nanoscale Science and Technology NanoFab.

\bibliography{report} 

\begin{thebibliography}{10}

\bibitem{Madau2014}
P.~Madau, M.~D., ``Cosmic star-formation history,'' {\em Annu. Rev. Astron.
  Astrophys.}~{\bf 52},  415--486 (2014).

\bibitem{Finkelstein2013}
S.~L.~Finkelstein, e.~a., ``A galaxy rapidly forming stars 700 million years
  after the big bang at redshift 7.51,'' {\em Annu. Rev. Astron.
  Astrophys.}~{\bf 502},  524–527 (2013).

\bibitem{Carilli2013}
C.~Carilli, F.~W., ``Cool gas in high-redshift galaxies,'' {\em Annu. Rev.
  Astron. Astrophys.}~{\bf 51},  105--161 (2013).

\bibitem{walter2020evolution}
Walter, F., Carilli, C., Neeleman, M., Decarli, R., Popping, G., Somerville,
  R.~S., Aravena, M., Bertoldi, F., Boogaard, L., Cox, P., et~al., ``The
  evolution of the baryons associated with galaxies averaged over cosmic time
  and space,'' {\em Astrophys. J.}~{\bf 902}(2),  111 (2020).

\bibitem{ade2020experiment}
Ade, P., Anderson, C., Barrentine, E., Bellis, N., Bolatto, A., Breysse, P.,
  Bulcha, B., Cataldo, G., Connors, J., Cursey, P., et~al., ``The experiment
  for cryogenic large-aperture intensity mapping {(EXCLAIM)},'' {\em J. Low
  Temp. Phys.}~{\bf 199},  1--11 (2020).

\bibitem{EXCLAIMSPIE}
Cataldo, G. et~al., ``Design and performance of a high resolution {$\mu$-Spec}:
  an integrated sub-millimeter spectrometer,'' in [{\em Millimeter,
  Submillimeter, and Far-Infrared Detectors and Instrumentation for Astronomy
  X}{\nolinebreak\hspace{0.1em}]},  International Society for Optics and
  Photonics (2020).

\bibitem{alam2015eleventh}
Alam, S., Albareti, F.~D., Prieto, C.~A., Anders, F., Anderson, S.~F.,
  Anderton, T., Andrews, B.~H., Armengaud, E., Aubourg, {\'E}., Bailey, S.,
  et~al., ``The eleventh and twelfth data releases of the sloan digital sky
  survey: final data from {SDSS-III},'' {\em Astrophys. J. Suppl. Ser.}~{\bf
  219}(1),  12 (2015).

\bibitem{barrentine2016design}
Barrentine, E.~M., Cataldo, G., Brown, A.~D., Ehsan, N., Noroozian, O.,
  Stevenson, T.~R., Kongpop, U.-Y., Wollack, E.~J., and Moseley, S.~H.,
  ``Design and performance of a high resolution {$\mu$-Spec}: an integrated
  sub-millimeter spectrometer,'' in [{\em Millimeter, Submillimeter, and
  Far-Infrared Detectors and Instrumentation for Astronomy
  VIII}{\nolinebreak\hspace{0.1em}]},   {\bf 9914},  99143O, International
  Society for Optics and Photonics (2016).

\bibitem{cataldo2015four}
Cataldo, G., Moseley, S.~H., and Wollack, E.~J., ``A four-pole power-combiner
  design for far-infrared and submillimeter spectroscopy,'' {\em Acta
  Astronaut.}~{\bf 114},  54--59 (2015).

\bibitem{cataldo2014micro}
Cataldo, G., Hsieh, W.-T., Huang, W.-C., Moseley, S.~H., Stevenson, T.~R., and
  Wollack, E.~J., ``Micro-spec: an ultracompact, high-sensitivity spectrometer
  for far-infrared and submillimeter astronomy,'' {\em Appl. Opt.}~{\bf 53}(6),
   1094--1102 (2014).

\bibitem{noroozian2015mu}
Noroozian, O., Barrentine, E., Brown, A., Cataldo, G., Ehsan, N., Hsieh, W.-T.,
  Stevenson, T., U-yen, K., Wollack, E., and Moseley, S.~H., ``$\mu$-spec: An
  efficient compact integrated spectrometer for submillimeter astrophysics,''
  in [{\em 26th International Symposium on Space Terahertz
  Technology}{\nolinebreak\hspace{0.1em}]},  (2015).

\bibitem{Yen}
Yen, H.~W., Friedrich, H.~R., Morrison, R.~J., and Tangonan, G.~L., ``{Planar
  Rowland spectrometer for fiber-optic wavelength demultiplexing},'' {\em Opt.
  Lett.}~{\bf 6 (12)},  639--641 (1981).

\bibitem{Bradford2004}
Bradford, C.~M., Ade, P. A.~R., Aguirre, J., Bock, J.~J., Duband, L., Earle,
  L., Glenn, J., Matsuhara, H., Naylor, B.~J., Nguyen, H., Yun, M., and
  Zmuidzinas, J., ``{Z-Spec: a broadband millimeter-wave grating spectrometer -
  design, construction, and first cryogenic measurements},'' in [{\em
  Millimeter and Submillimeter Detectors for Astronomy
  II}{\nolinebreak\hspace{0.1em}]},  {\em Society of Photo-Optical
  Instrumentation Engineers (SPIE)} {\bf 5408},  257--267 (2004).

\bibitem{Shirokoff}
Shirokoff, E., Barry, P.~S., Bradford, C.~M., Chattopadhyay, G., Day, P.,
  Doyle, S., Hailey-Dunsheath, S., Hollister, M.~I., Kovács, A., Leduc, H.~G.,
  McKenney, C., Mauskopf, P., O'Brient, R., Padin, S., Reck, T., Swenson,
  L.~J., Tucker, C.~E., and Zmuidzinas, J., ``{Design and Performance of
  SuperSpec: An On-Chip, KID-based, mm-Wavelength Spectrometer},'' {\em Journal
  of Low Temperature Physics}~{\bf 176}(5-6),  657--662 (2014).

\bibitem{Endo}
Endo, A., Sfiligoj, C., Yates, S. J.~C., Baselmans, J. J.~A., Thoen, D.~J.,
  Javadzadeh, S. M.~H., van der Werf, P.~P., Baryshev, A.~M., and Klapwijk,
  T.~M., ``{On-chip filter bank spectroscopy at 600-700 GHz using NbTiN
  superconducting resonators},'' {\em Applied Physics Letters}~{\bf 103},
  032601 (2013).

\bibitem{datta2013large}
Datta, R., Munson, C., Niemack, M., McMahon, J., Britton, J., Wollack, E.~J.,
  Beall, J., Devlin, M., Fowler, J., Gallardo, P., et~al., ``Large-aperture
  wide-bandwidth antireflection-coated silicon lenses for millimeter
  wavelengths,'' {\em Appl. Opt.}~{\bf 52}(36),  8747--8758 (2013).

\bibitem{loewenstein1973optical}
Loewenstein, E.~V., Smith, D.~R., and Morgan, R.~L., ``Optical constants of far
  infrared materials. 2: Crystalline solids,'' {\em Appl. Opt.}~{\bf 12}(2),
  398--406 (1973).

\bibitem{afsar1994millimeter}
Afsar, M.~N. and Chi, H., ``Millimeter wave complex refractive index, complex
  dielectric permittivity and loss tangent of extra high purity and compensated
  silicon,'' {\em Int. J. Infrared Millim. Waves}~{\bf 15}(7),  1181--1188
  (1994).

\bibitem{wollack2020infrared}
Wollack, E.~J., Cataldo, G., Miller, K.~H., and Quijada, M.~A., ``Infrared
  properties of high-purity silicon,'' {\em Opt. Lett.}~{\bf 45}(17),
  4935--4938 (2020).

\bibitem{iacono2011line}
Iacono, A., Freni, A., Neto, A., and Gerini, G., ``In-line x-slot element focal
  plane array of kinetic inductance detectors,'' in [{\em Proceedings of the
  5th European Conference on Antennas and Propagation
  (EUCAP)}{\nolinebreak\hspace{0.1em}]},   3316--3320, IEEE (2011).

\bibitem{EXCLAIMopticsSPIE}
Essinger-Hileman, T. et~al., ``Optical design of the experiment for cryogenic
  large-aperture intensity mapping {(EXCLAIM)},'' in [{\em Millimeter,
  Submillimeter, and Far-Infrared Detectors and Instrumentation for Astronomy
  X}{\nolinebreak\hspace{0.1em}]},  International Society for Optics and
  Photonics (2020).

\bibitem{rotman1963wide}
Rotman, W. and Turner, R., ``Wide-angle microwave lens for line source
  applications,'' {\em IEEE Trans. Antennas Propag.}~{\bf 11}(6),  623--632
  (1963).

\bibitem{rowland1883xxix}
Rowland, H.~A., ``{XXIX.} on concave gratings for optical purposes,'' {\em
  Philos. Mag.}~{\bf 16}(99),  197--210 (1883).

\bibitem{microspec3}
Cataldo, G., Barrentine, E., Bulcha, B., Ehsan, N., Hess, L., Noroozian, O.,
  Stevenson, T., Wollack, E., Moseley, S., and Switzer, E., ``Second-generation
  {Micro-Spec}: A compact spectrometer for far-infrared and submillimeter space
  missions,'' {\em Acta Astronaut.}~{\bf 162},  155--157 (2019).

\bibitem{baselmans2017kilo}
Baselmans, J.~J., Bueno, J., Yates, S.~J., Yurduseven, O., Llombart, N.,
  Karatsu, K., Baryshev, A., Ferrari, L., Endo, A., Thoen, D., et~al., ``A
  kilo-pixel imaging system for future space based far-infrared observatories
  using microwave kinetic inductance detectors,'' {\em Astron. Astrophys.}~{\bf
  601},  A89 (2017).

\bibitem{bulcha2018electromagnetic}
Bulcha, B., Cataldo, G., Stevenson, T., U-Yen, K., Moseley, S., and Wollack,
  E., ``Electromagnetic design of a magnetically coupled spatial power
  combiner,'' {\em J. Low Temp. Phys.}~{\bf 193}(5-6),  777--785 (2018).

\bibitem{wollack2014impedance}
Wollack, E., Chuss, D., Rostem, K., and U-Yen, K., ``Impedance matched
  absorptive thermal blocking filters,'' {\em Rev. Sci. Instrum.}~{\bf 85}(3),
  034702 (2014).

\bibitem{patel2013fabrication}
Patel, A., Brown, A.-D., Hsieh, W.-T., Stevenson, T., Moseley, S.~H., Kongpop,
  U., Ehsan, N., Barrentine, E., Manos, G., Wollack, E.~J., et~al.,
  ``Fabrication of {MKIDS} for the {MicroSpec} spectrometer,'' {\em IEEE Trans.
  Appl. Supercond.}~{\bf 23}(3),  2400404--2400404 (2013).

\bibitem{brown2018fabrication}
Brown, A.-D., Brekosky, R., Franz, D., Hsieh, W.-T., Kutyrev, A., Mikula, V.,
  Miller, T., Moseley, S.~H., Oxborrow, J., Rostem, K., et~al., ``Fabrication
  of ultrasensitive {TES} bolometric detectors for {HIRMES},'' {\em J. Low
  Temp. Phys.}~{\bf 193}(5-6),  675--680 (2018).

\bibitem{brown2015high}
Brown, A.~D. and Patel, A.~A., ``High precision metal thin film liftoff
  technique,'' (July~7 2015).
\newblock US Patent 9,076,658.

\bibitem{stevenson2009superconducting}
Stevenson, T.~R., Adams, J.~S., Hsieh, W.-T., Moseley, S.~H., Travers, D.~E.,
  Kongpop, U., Wollack, E.~J., Zmuidzinas, J., et~al., ``Superconducting films
  for absorber-coupled mkid detectors for sub-millimeter and far-infrared
  astronomy,'' {\em IEEE Trans. Appl. Supercond.}~{\bf 19}(3),  561--564
  (2009).

\bibitem{brown2016impact}
Brown, A.~D., Barrentine, E.~M., Moseley, S.~H., Noroozian, O., and Stevenson,
  T., ``The impact of standard semiconductor fabrication processes on
  polycrystalline {Nb} thin-film surfaces,'' {\em IEEE Trans. Appl.
  Supercond.}~{\bf 27}(4),  1--5 (2016).

\end{thebibliography}
\bibliographystyle{spiebib} 

\end{document}